*Краткие сообщения*

# On the determination of the quasiparticle scattering rate in unconventional superconductors by microwave surface impedance

N.T. Cherpak[1], A.A. Barannik[1], R. Prozorov[2,3], M. Tanatar[2], and A.V. Velichko[1]

[1]*A. Usikov Institute of Radiophysics and Electronics of the National Academy of Sciences of Ukraine 12 Acad. Proskura Str., Kharkiv 61085, Ukraine*

[2]*Department of Physics and Astronomy, Iowa State University, Ames, IA 50011, USA*

[3]*Ames Laboratory USDOE, Ames, IA 50011, USA*
E-mail: cherpak@ire.kharkov

As found from numerous microwave experiments on the unconventional Fe-based superconductors, the temperature dependence of the quasiparticle scattering rate $\tau^{-1}$ cannot be accurately described within the framework of standard Drude module in the popular approximation of $\omega\tau \ll 1$, where $\omega$ is the signal frequency. To account for the discrepancy, we have extended the classical Drude model for the case of arbitrary values of $\omega\tau$, and obtained the expression for $\tau^{-1}$ as a function of experimentally measurable quantities, namely the real and imaginary parts of the microwave surface impedance. We then show the temperature dependence of $\tau^{-1}$ in superconducting Ba(Fe$_{1-x}$Co$_x$)$_2$As$_2$ single crystal pnictide derived from the Ka-band surface impedance measurements within the framework of the modified expression. The measurements indicate the extent to which assumption of $\omega\tau \ll 1$ gives results different from those obtained without this restriction, i.e., incorrect results.

PACS: *74.25.nn* Surface impedance;
74.70.Xa Pnictides and chalcogenides.

Keywords: quasiparticle conductivity, microwave surface impedance, Fe-pnictide.

Studying the temperature dependence of the quasiparticle scattering rate $\tau^{-1}$ offers a great insight into the underlying physics of superconductors. Here, the absolute value and the temperature dependence of $\tau^{-1}$ reflect very important characteristics of the electron system of the materials [1]. Of particular interest is studying the electron system in the unconventional superconductors (in particular the high-$T_c$ cuprates and Fe-based superconductors), and recently a plethora of various experimental techniques covering a wide range of signal frequencies have been used to accomplish this purpose (see, e.g., [2–5]). Superconducting single crystals and films were measured using dc signal by, e.g., magnetic-force microscopy and scanning SQUID [6,7]. Measurements by means of radio-frequency tunnel-diode resonators [8–10], microwave-range resonance cavities [11–15] as well as THz and optical reflectivity [16,17] techniques have also been reported.

As far as microwave measurements are concerned, once can use surface impedance data to extract complex conductivity of the superconductor which, in case, gives us an opportunity to extract the temperature-dependent London penetration depth $\lambda_L$ and the quasiparticle scattering rate $\tau^{-1}$. In addition, since the scattering in the normal state is directly related to the superconducting pairing strength [18], extension of those measurements into a superconducting state is of notable interest.

Apparently, microwave and higher frequency measurements are the only kind of experiments that allow one to determine the complex conductivity in a superconducting state [19], which in turn can be used to extract $\tau^{-1}$ and the interesting fact is that the values of $\tau$ have been found to dramatically increase in the unconventional superconductors (see, e.g., [20]).

The task of finding $\tau$ is usually straightforward in the case of $\omega\tau \ll 1$, however in the unconventional superconductors where $\tau^{-1}$ is strongly temperature-dependent [11–14], and as the signal frequencies $\omega$ increase towards the millimeter wave range [15], researches often come across the difficulty of processing the experimental data for the case of arbitrary $\tau$ values.

In this paper we address the problem of obtaining the generalized expression for the quasiparticle scattering rate in terms of microwave surface impedance valid for arbitrary values of $\omega\tau$, and then extract the temperature dependence of $\tau^{-1}$ in the experimentally measured single crystal of BaFeCoAs.

Within framework of the local electrodynamics, microwave surface impedance of the conducting materials is determined as (in SI) [21]

$$Z_s = \sqrt{\frac{i\omega\mu_0}{\sigma}} = R_s + iX_s, \qquad (1)$$

where $\omega$ is the frequency of electromagnetic field, which varies as $e^{i\omega t}$; $\mu_0$ is the magnetic permeability of vacuum; $\sigma$ is the microwave conductivity, which is a complex value $\sigma = \sigma' - i\sigma''$; $R_s$ and $X_s$ are surface resistance and reactance of the conductor. Using different techniques of measurements in the microwave range, one can accurately determine experimentally values of $R_s$ and $X_s$ (see, e.g., [19,22]), which in turn determine the conductivity $\sigma$.

According to the two-fluid model there are two currents: a superconducting current, conditioned by superfluid component, and the normal current attributed to quasiparticles. Correspondingly, the conductivity $\sigma$ in (1) can be written as [23]

$$\sigma = \sigma_s + \sigma_n = \frac{e^2}{m}\left[\frac{n_s}{i\omega} + \frac{n_n\tau}{1+i\omega\tau}\right] = \sigma' - i\sigma'', \qquad (2)$$

where $\sigma' = \sigma'_l - i\sigma''_l$, $n_n$ and $n_s$ are electron concentration of quasiparticle and superfluid components accordingly, $n_n + n_s = n$, where $n$ is constant; $e$ and $m$ are the charge and mass of electrons. The conductivity $\sigma_n$ is written in the assumption of validity of the Drude model.

We now need to express the conductivity (2) in terms of $R_s(T)$ and $X_s(T)$ because these are the quantities measured in the experiments:

$$\sigma' = 2\omega\mu_0 \frac{X_s R_s}{|Z_s|^4}, \qquad \sigma'' = \omega\mu_0 \frac{X_s^2 - R_s^2}{|Z_s|^4}, \qquad (3)$$

where $|Z_s|^4 = (X_s^2 + R_s^2)^2$.

It is worth noting that at $\omega\tau \ll 1$ the conductivity can be written in the form

$$\sigma = \frac{e^2}{m}\left[n_n\tau - i\frac{n_s}{\omega}\right] = \sigma_1 - i\sigma_2; \quad \sigma''_1 = 0; \quad \sigma_2 = i\sigma_s. \qquad (4)$$

At arbitrary values of $\omega\tau$, on the other hand, from (2) and (4) we obtain

$$\sigma' = \sigma'_1 = \frac{\sigma_1}{1+\omega^2\tau^2}; \quad \sigma'' = \sigma_2 + \sigma''_1 = \sigma_2 + \frac{\sigma_2\omega\tau}{1+\omega^2\tau^2}, \qquad (5)$$

where $\sigma_2 = \frac{1}{\omega\mu_0\lambda_L^2(T)}$.

On the assumption that in the superconducting state at sufficiently low temperatures $n_n(0) = 0$ and $n_s(0) = n$, one can use equations (5) to obtain [23]

$$\frac{1}{\tau} = \omega\left[\frac{\sigma''(0)}{\sigma'(T)} - \frac{\sigma''(T)}{\sigma'(T)}\right]; \quad \sigma''(0) = \sigma_2(0) = \frac{1}{\lambda_L^2(0)\mu_0\omega}. \qquad (6)$$

Substituting (3) into (6) yields

$$\frac{1}{\tau} = \frac{1}{\mu_0\sigma'_1(T)\lambda_L^2(0)} - \omega\frac{X_s^2 - R_s^2}{2X_s R_s}. \qquad (7)$$

When $\omega\tau \ll 1$ the expression (7) becomes the well known formula [24]

$$\frac{1}{\tau} = \frac{1-[\lambda_L^2(0)/\lambda_L^2(T)]}{\mu_0\sigma'_1(T)\lambda_L^2(0)}. \qquad (8)$$

Figure 1 shows the scattering rate $\tau^{-1}$ in the optimally doped pnictide single crystal Ba(Fe$_{1-x}$Co$_x$)$_2$As$_2$ [8,9] as a function of temperature. The data are obtained from Ka-band microwave impedance measurements by using high Q-factor quasioptical slotted sapphire resonator (excited in whispering gallery modes) with YBCO end plates, and the expressions (7) and (8) are used to process the impedance measurement data.

It is seen that (7) gives a significant correction in $\tau^{-1}(T)$ at low temperatures for $\omega\tau$ exceeding 0.05. In turn, a more accurate estimate of $\tau$ gives an opportunity to more accurately obtain $\sigma_2(T)$, which is determined by the measured $R_s(T)$ and $X_s(T)$ and offers information about the penetration depth $\lambda_L(T)$ and the structure of the superconducting energy gap [3].

Finally, we should emphasize that the generalized expression for $\tau$ (Eq. (7)) derived in this work and valid for arbitrary values of $\omega\tau$ in combination with the novel resonator technique with HTS end plates [22,25] enabled us to study the temperature dependence of the quasiparticle scattering rate $\tau^{-1}$ for the whole family of the Fe-based superconductors including Fe-pnictides [25], Fe-chalcogenides [14] and others at microwave frequencies.

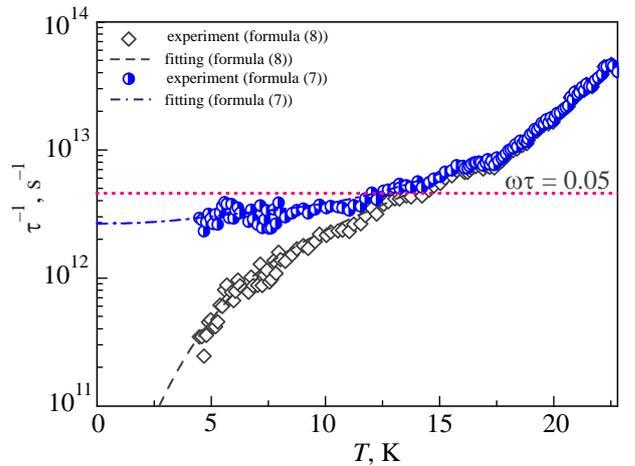

*Fig. 1.* The temperature dependence of the quasiparticle scattering rate $\tau^{-1}$ in a single crystal of optimally-doped Ba(Fe$_{1-x}$Co$_x$)$_2$As$_2$ calculated using the generalized expressions (7) and (8) valid at $\omega\tau \ll 1$. The horizontal dotted line shows $\tau^{-1}$ for $\omega\tau = 0.05$.

The authors thank Dr. S.A. Vitusevich, FZ Jülich Peter Grünberg Institute, Germany, for supporting this research.

1. D.A. Bonn, S. Kamal, Kuan Zhang, Ruixing Liang, D.J. Baar, E. Klein, and W.N. Hardy, *Phys. Rev. B* **50**, 4051 (1994).
2. P.J. Hirschfeld, M.M. Korshunov, and I.I. Mazin, *Rep. Prog. Phys.* **74**, 124508 (2011).
3. R. Prozorov and V.G. Kogan, *Rep. Prog. Phys.* **74**, 124505 (2011).
4. G.R. Stewart, *Rev. Mod. Phys.* **83**, 1589 (2011).
5. A. Chubukov, *Annu. Rev. Condens. Matter Phys.* **3**, 57 (2012).
6. L. Luan, O.M. Auslaender, T.M. Lippman, C.W. Hicks, B. Kalisky, J.-H. Chu, J.G. Analytis, I.R. Fisher, J.R. Kirtley, and K.A. Moler, *Phys. Rev. B* **81**, 100501 (2010).
7. L. Luan, T.M. Lippman, C.W. Hicks, J.A. Bert, O.M. Auslaender, J.-H. Chu, J.G. Analytis, I.R. Fisher, and K.A. Moler, *Phys. Rev. Lett.* **106**, 067001 (2011).
8. R.T. Gordon, N. Ni, C. Martin, M.A. Tanatar, M.D. Vannette, H. Kim, G.D. Samolyuk, J. Schmalian, S. Nandi, A. Kreyssig, A.I. Goldman, J.Q. Yan, S.L. Bud'ko, P.C. Canfield, and R. Prozorov, *Phys. Rev. Lett.* **102**, 127004 (2009).
9. R.T. Gordon, C. Martin, H. Kim, N. Ni, M.A. Tanatar, J. Schmalian, I.I. Mazin, S.L. Budko, P.C. Canfield, and R. Prozorov, *Phys. Rev. B* **79**, 100506(R) (2009).
10. H. Kim, R.T. Gordon, M.A. Tanatar, J. Hua, U. Welp, W.K. Kwok, N. Ni, S.L. Budko, P.C. Canfield, A.B. Vorontsov, and R. Prozorov, *Phys. Rev. B* **82**, 060518 (2010).
11. K. Hashimoto, T. Shibauchi, T. Kato, K. Ikada, R. Okazaki, H. Shishido, M. Ishikado, H. Kito, A. Iyo, H. Eisaki, S. Shamoto, and Y. Matsuda, *Phys. Rev. Lett.* **102**, 017002 (2009).
12. K. Hashimoto, T. Shibauchi, S. Kasahara, K. Ikada, S. Tonegawa, T. Kato, R. Okazaki, C.J. van der Beek, M. Konczykowski, H. Takeya, K. Hirata, T. Terashima, and Y. Matsuda, *Phys. Rev. Lett.* **102**, 207001 (2009).
13. J.S. Bobowski, J.C. Baglo, J. Day, P. Dosanjh, R. Ofer, B.J. Ramshaw, R. Liang, D.A. Bonn, W.N. Hardy, H. Luo, Z.-S. Wang, L. Fang, and H.-H. Wen, *Phys. Rev. B* **82**, 094520 (2010).
14. H. Takahashi, Y. Imai, S. Komiya, I. Tsukada, and A. Maeda, *Phys. Rev. B* **84**, 132503 (2011).
15. A.A. Barannik, N.T. Cherpak, N. Ni, M.A. Tanatar, S.A. Vitusevich, V.N. Skresanov, P.C. Canfield, R. Prozorov, V.V. Glamazdin, and K.I. Torokhtii, *Fiz. Nizk. Temp.* **37**, 912 (2011) [*Low Temp. Phys.* **37**, 725 (2011)].
16. R. Valdes Aguilar, L.S. Bilbro, S. Lee, C.W. Bark, J. Jiang, J.D. Weiss, E.E. Hellstrom, D.C. Larbalestier, C.B. Eom, and N.P. Armitage, *Phys. Rev. B* **82**, 180514 (2010).
17. T. Fischer, A.V. Pronin, J. Wosnitza, K. Iida, F. Kurth, S. Haindl, L. Schultz, B. Holzapfel, and E. Schachinger, *Phys. Rev. B* **82**, 224507 (2010).
18. L. Taillefer, *Annu. Rev. Condens. Matter Phys.* **1**, 51 (2010).
19. D.A. Bonn and W.N. Hardy, Chapter 2, *Microwave Surface Impedance of High Superconductors*, In a book *Physical Properties of High Temperature Superconductors*, M. Ginsberg (ed.), World Scientific (1996).
20. R. Harris, P.J. Turner, Saeid Kamal, A.R. Hosseini, P. Dosanjh, G.K. Mullins, J.S. Bobowski, C.P. Bidinosti, D.M. Broun, Ruixing Liang, W.N. Hardy, and D.A. Bonn, *Phys. Rev. B* **74**, 104508 (2006).
21. L.D. Landau and E.M. Lifshitz, *Electrodynamics of Continuous Media*, Vol. 8, Pergamon Press, Oxford (1960).
22. N.T. Cherpak, A.A. Barannik, S.A. Bunyaev, Y.V. Prokopenko, K.I. Torokhtii, and S.A. Vitusevich, *IEEE Trans. on Appl. Supercond.* **21**, 591 (2011).
23. J.R. Waldram, P. Theopistou, A. Porch, and H.-M. Cheah, *Phys. Rev. B* **55**, 3222 (1997).
24. E. Schachinger and J.P. Carbotte, *Phys. Rev. B* **80**, 174526 (2009).
25. A. Barannik, N.T. Cherpak, M.A. Tanatar, S. Vitusevich, V. Skresanov, P.C. Canfield, and R. Prozorov, *Phys. Rev. B* (2013), to be published.